\documentclass[aps,prl,reprint,superscriptaddress,nofootinbib,longbibliography,floatfix]{revtex4-2}

\usepackage[T1]{fontenc}
\usepackage[utf8]{inputenc}
\usepackage{lmodern}
\usepackage{amsmath,amssymb,bm,mathtools}
\usepackage{graphicx}
\usepackage{booktabs}
\usepackage{array}
\usepackage{microtype}
\usepackage{xcolor}
\usepackage{hyperref}
\usepackage{enumitem}
\usepackage{physics}
\usepackage{comment}
\hypersetup{colorlinks=true,linkcolor=blue!55!black,citecolor=blue!55!black,urlcolor=blue!55!black}

\newcommand{\workbeta}{6}
\newcommand{\workJzero}{1}
\newcommand{\criticalDelta}{0.44}
\newcommand{\criticalBand}{0.43\text{--}0.45}

\newcommand{\SUtwo}{\mathrm{SU}(2)}
\newcommand{\EN}{E_N}
\newcommand{\FN}{F_N}
\newcommand{\Veff}{\mathcal V_N}
\newcommand{\Nadj}{N_{\mathrm{adj}}}
\newcommand{\avgT}[1]{\left\langle #1\right\rangle_T}
\newcommand{\avgJ}[1]{\left[#1\right]_J}

\begin{document}
\title{Frustration without Glass: A Non-Abelian Gauge Model of Network Compatibility}

\author{Xuanhua Wang}
\affiliation{School of Arts and Sciences, Fuyao University of Science and Technology, Fuzhou, Fujian 350122, China}

\begin{abstract}

We formulate a non-Abelian theory of network compatibility in which dynamical transformations reside on the links. Gauge covariance follows from the freedom to choose local representation frames, while plaquette holonomies quantify the incompatibility of closed-loop transformations. For an $\SUtwo$ model on the complete simplicial $2$-complex with quenched random plaquette couplings, parallel-tempering simulations reveal a continuous disorder-driven phase transition characterized by the network compatibility $M_P$. As the disorder strength approaches the critical value, the compatibility $M_P$ drops rapidly to a value that decreases with system size, while the frustration energy of the network sharply rises. Moreover, analysis of connected replica-overlap width provides no evidence for thermodynamic replica-symmetry breaking. Instead, the high-disorder regime sustains a nearly constant integrated adjacent correlation $\mathcal I_{\mathrm{adj}}$ as the system size increases. Therefore, rather than a frozen gauge glass or a featureless disordered ``gas'' phase, dense topological frustration produces a non-glassy correlated gauge liquid in which individual pair correlations are geometrically diluted while a finite integrated correlation survives.

\end{abstract}

\keywords{gauge theory, structural balance, higher-order networks, simplicial complex, quenched disorder, correlated liquid, gauge glass, phase transition}
\maketitle

Most statistical models of collective opinion assign dynamical variables to network vertices, while links primarily specify adjacency or interaction strength \cite{castellano2009statistical,Starnini2026Opinion,HolmeNewman2006,Baumann2021Multidimensional}. Ising- and voter-type models employ discrete states, whereas bounded-confidence models represent opinions by scalar or vector coordinates \cite{castellano2009statistical,Starnini2026Opinion,HegselmannKrause2002}. In either case, the interaction network generally acts as a substrate for node dynamics and communication channels do not transform the information being transmitted. In realistic communication, however, channels are rarely neutral; information undergoes systematic transformations—including reframing, selective filtering, and translation—during transmission. Recent developments have begun to address this by introducing state-dependent edge operations, such as discourse sheaves and topological-synchronization dynamics on higher-dimensional simplices \cite{HansenGhrist2021,Tian2025Matrix,Millan2020,Ghorbanchian2021,Carletti2023}. The statistical mechanics of such networks, nevertheless, remains poorly understood. 
It is unclear whether they can support any forms of thermodynamic order or phase transitions.

In this Letter, we formulate a statistical framework in which the dynamical variables reside directly on the links. Every node describes a multidimensional issue in a local frame, and an oriented relation link carries a transformation $U_{ij}\in G$ that maps information represented in the frame of $j$ to that of $i$. Because the choice of local frame is arbitrary and may be changed independently at every node, physical observables must remain invariant under local frame redefinitions $h_i \in G$, enforcing the transformation law $U_{ij}\longrightarrow h_iU_{ij}h_j^{-1}$. Gauge covariance thus follows from the local representational freedom of the network. The physically meaningful quantities are consequently gauge-invariant combinations of link transformations, which compare the accumulated maps along alternative paths.

The choice of $G$ is determined by how relational transformations compose. Communication and judgment can be sequence dependent: if two transformations $\mathcal T_A$ and $\mathcal T_B$ obey $\mathcal T_B\mathcal T_A\neq\mathcal T_A\mathcal T_B$, reversing their order changes the resulting representation. Abelian groups such as $\mathrm{U}(1)$ cannot encode this distinction, and we use $\SUtwo$ as a minimal compact non-Abelian prototype.  For $G=\mathbb Z_2$, loop consistency reduces to structural balance on signed networks \cite{Antal2005,Marvel2009,Gorski2020}. The ordered product of link maps along a path gives the accumulated transformation; two alternative paths are compatible only when the holonomy around the closed loop joining them is trivial \cite{Gao2021Geometry,Wegner1971,Wilson1974,Kogut1979}. This formulation generalizes consensus from equality of node states to \emph{global compatibility of network relations}.

The central question is which collective phase replaces global compatibility when quenched loop preferences become mutually frustrated. In conventional disordered systems, dense quenched frustration can produce glassy locking \cite{EdwardsAnderson1975,SherringtonKirkpatrick1975,BinderYoung1986,KosterlitzAkino1998}; whether the same occurs for geometrically constrained non-Abelian holonomies is not evident. We address this question using fluctuating $\SUtwo$ relations on the complete simplicial $2$-complex with quenched random plaquette couplings. A gauge-invariant decomposition separates uniform compatibility, disorder-pinned local structure, connected correlations between distinct loops, and replica locking. Increasing disorder drives a continuous loss of global compatibility into a correlated gauge liquid: the integrated adjacent correlation weight
remains finite, while individual pair correlations are geometrically diluted and the connected replica-overlap distribution collapses toward zero. Dense frustration therefore produces neither independent local disorder nor a thermodynamic gauge glass.

\textit{Non-Abelian gauge network.}--
We represent a higher-order communication network by an oriented simplicial $2$-complex $\mathcal{K}=(V,E,F)$ \cite{Millan2020,Ghorbanchian2021,Carletti2023}. Its vertices represent agents or communities, its edges represent pairwise communication channels, and its faces are formed by triangular plaquettes. Each oriented edge $(i,j)$ carries a relational transformation or connection $U_{ij}\in\mathrm{SU}(2)$, and $U_{ji}=U_{ij}^{-1}=U_{ij}^{\dagger}$. Under independent changes of the local frames, $h_i\in\mathrm{SU}(2)$, the edge variables transform as $U_{ij}\longrightarrow h_iU_{ij}h_j^{-1}$. For an oriented triangular face $f=(i,j,k)$, the accumulated relational transformation is
\begin{equation}
U_f=U_{ij}U_{jk}U_{ki}, \quad P_f=\frac{1}{2}\operatorname{ReTr}U_f,
\label{eq:plaquette_definition}
\end{equation}
where the plaquette holonomy $U_f$ encodes the local curvature of the connection around $f$, and the Wilson plaquette observable $P_f\in [-1,1]$  is gauge invariant and measures the local path compatibility. The limit $P_f=1$ corresponds to exact compatibility of the relational transformations around the triangle. The plaquette holonomies are generated by shared links and are not themselves independent degrees of freedom. For four distinct vertices $i,j,k,\ell$, they obey non-Abelian compatibility relations
\begin{equation}
U_{ijk}U_{ik\ell}U_{i\ell j}
=U_{ij}U_{jk\ell}U_{ji}.
\label{eq:tetrahedral_identity}
\end{equation}
This discrete Bianchi-type identity relates the holonomies of the four faces of a tetrahedron and is the simplicial analogue of lattice Bianchi constraints \cite{Kogut1979}.

We assign a quenched coupling $J_f$ to each triangular unit and consider the Hamiltonian
\begin{equation}
\mathcal{H}_J[U]
=-\sum_{f\in F}J_f\operatorname{ReTr}U_f
=-2\sum_{f\in F}J_fP_f.
\label{eq:hamiltonian}
\end{equation}
A positive $J_f$ favors relational closure, whereas a negative $J_f$ favors a strongly nontrivial holonomy. The couplings thus represent persistent heterogeneity in the consistency pressures acting on different triangular units. The partition function at statistical temperature $\beta^{-1}$ is $Z_J=\int\prod_{(i,j)\in E}\mathrm{d}U_{ij}\,\exp\left(-\beta \mathcal{H}_J\right)$, where $\mathrm{d}U$ is the normalized Haar measure on $\mathrm{SU}(2)$. 

We specialize to the complete simplicial $2$-complex on $N$ vertices with $E_N=\binom{N}{2}$ edges and $F_N=\binom{N}{3}$ faces.  Each edge belongs to $c_N=N-2$ triangular units. To obtain a nontrivial dense-network limit, we use Sherrington-Kirkpatrick (SK) normalization and draw the couplings independently from
\begin{equation}
J_f=\frac{J_0}{c_N}+ \frac{\Delta}{\sqrt{c_N}}\xi_f, \quad \xi_f\sim\mathcal{N}(0,1),
\label{eq:disorder_distribution}
\end{equation}
where the $\xi_f$ are Gaussian variables with zero mean and unit variance. This is the dense-coupling scaling of the SK model adapted to the face coordination $c_N$ of an edge \cite{SherringtonKirkpatrick1975}. The coherent contribution to the local field acting on an edge scales as $c_N\frac{J_0}{c_N}=J_0$, whereas the root-mean-square random contribution scales as $\sqrt{c_N}\frac{\Delta}{\sqrt{c_N}}=\Delta$. Both remain finite as $N\rightarrow\infty$. For uniform positive $J_f$, the partition function reduces to a plaquette gauge action; quenched face dependence and dense-network scaling define a random-coupling simplicial generalization \cite{Wegner1971,Wilson1974,Kogut1979,WangHarringtonPreskill2003}. The quenched randomness of $J_f$ introduces frustrated plaquette configurations. As shown in the Supplemental Material (SM), the resulting frustration energy—the excess energy per link originating from mutually incompatible local energy minima—grows rapidly with the disorder strength \cite{sm}.

The complete complex provides a controlled dense limit rather than a necessity of the framework. It separates the effects of relational frustration from those of finite connectivity: every link participates in $c_N=N-2$ triangular units, the number of gauge variables grows as $E_N=O(N^2)$, and the number of loop constraints grows as $F_N=O(N^3)$. The SK scaling keeps the net field on each link finite and makes the free energy extensive in $E_N$. 

Since local gauge symmetry precludes a gauge-variant link order parameter \cite{Elitzur1975}, we monitor the global compatibility through the gauge-invariant average
\begin{equation}
m_P = \frac{1}{|F|}\sum_{f \in F} P_f, \quad
M_P = \left[ \left\langle m_P \right\rangle_T \right]_J,
\label{eq:plaquette_order_parameter}
\end{equation}
where $\langle \cdots \rangle_T$ denotes the thermal average at a fixed disorder realization, and $[\cdots]_J$ denotes the disorder average. In the network, quenched disorder competes with the globally coherent plaquette pattern. However, the geometric constraints in Eq.~\eqref{eq:tetrahedral_identity} prevent the triangular units from responding independently, collectively giving rise to distinct collective patterns as disorder varies.

\textit{Spatial correlations.---}
For a fixed disorder realization, define the equilibrium profile $\mu_f^{(J)}=\avgT{P_f}$, the fluctuation $\delta P_f=P_f-\mu_f^{(J)}$, and the connected covariance
\begin{equation}
 G_{fg}^{(J)}=\avgT{\delta P_f\delta P_g}.
 \label{eq:connected_covariance}
\end{equation}
The conventional replica overlap contains the squared one-point profile and can remain nonzero even for independent replicas; the exact decomposition is given in End Matter. To quantify the correlation carried by adjacent simplices, we introduce the integrated adjacent correlation weight
\begin{equation}
 \mathcal I_{\mathrm{adj}}=\avgJ{\frac1{\FN}\sum_f\sum_{g\sim f} \left(G_{fg}^{(J)}\right)^2},
 \label{eq:Iadj}
\end{equation}
where $g\sim f$ denotes a distinct triangular face sharing an edge with $f$. Thus $\mathcal I_{\mathrm{adj}}$ is the average total squared connected covariance carried by the adjacency neighborhood of one plaquette. 

\begin{figure*}[t]
\centering
\includegraphics[width=\textwidth]{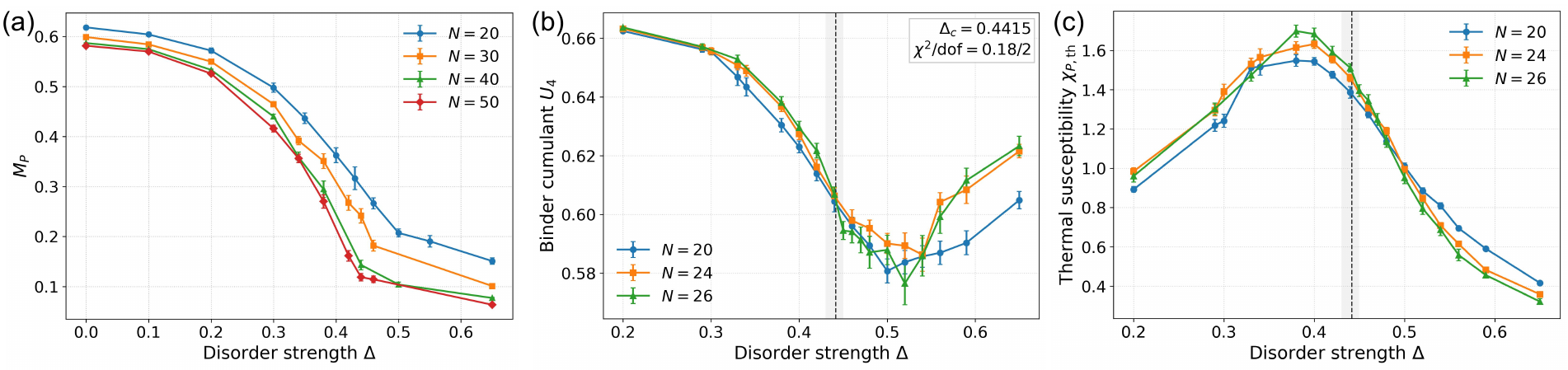}
\caption{Continuous disorder-driven loss of global compatibility at $J_0=1$ and $\beta=6$. (a) Gauge-invariant order parameter $M_P$ as a function of disorder strength. 
(b) Binder cumulant $U_4$, showing a common crossing within the window $\Delta_c\simeq0.44 \pm0.01$. (c) Connected thermal susceptibility of the plaquettes $\chi_{P,\mathrm{con}}$, whose maximum grows and peak width shrinks with the system size. Error bars represent disorder-resolved statistical uncertainties.}
\label{fig:thermodynamics}
\end{figure*}

For two independent replicas $\alpha$ and $\gamma$ at the same disorder, the connected replica overlap and its width are defined as
\begin{equation}
 q_{\alpha\gamma}^{\mathrm c}
 =\frac1{\FN}\sum_f\delta P_f^\alpha\delta P_f^\gamma, \quad  W_{\mathrm c}=\avgJ{\avgT{(q_{\alpha\gamma}^{\mathrm c})^2}}.
 \label{eq:connected_overlap}
\end{equation}
The connected Parisi overlap width can be rewritten as $W_{\mathrm c} =\avgJ{\mathrm{Var}_T(q_{\alpha\beta}^c)}=\frac1{\FN^2}\avgJ{\sum_{f,g}\left(G_{fg}^{(J)}\right)^2}$. Consequently, $\mathcal I_{\mathrm{adj}}$ is precisely the adjacent-sector contribution to $\FN W_{\mathrm c}$. This separates a finite integrated correlation from thermodynamic replica locking, which requires a nontrivial overlap sector to persist as the system grows \cite{Parisi1983,BinderYoung1986}.

\emph{Continuous phase transition.---}
We investigate the model using local Metropolis updates \cite{Metropolis1953} combined with parallel tempering \cite{HukushimaNemoto1996}. We set $J_0=1$ and focus on the low-temperature regime $\beta=6$, where the weak-disorder system is close to a coherent phase. We introduce the Binder cumulant $U_4=1-\frac{\left[\left\langle m_P^4\right\rangle_T\right]_J}{3\left[\left\langle m_P^2\right\rangle_T\right]_J^2},$ and the connected plaquette susceptibility $\chi_{\mathrm{con}}=\beta E_N\left[\left\langle m_P^2\right\rangle_T-\left\langle m_P\right\rangle_T^2\right]_J$ to characterize the transition \cite{FisherBarber1972,Binder1981}. The normalization by $E_N$ follows from the extensive scaling of the free energy with the number of edges.

Figure~\ref{fig:thermodynamics}(a) shows that $M_P$ is $\mathcal O(1)$ at weak disorder and decreases rapidly across a size-dependent transition region. The Binder cumulants meet within the common interval $\Delta=[0.43,0.45]$, giving the reference estimate $\Delta_c\simeq0.44$ [Fig.~\ref{fig:thermodynamics}(b)]. This is consistent with the $M_P$ curves and is compatible with the small $\chi^2$ value, which supports a unique critical crossing. At the same time, $\chi_{\mathrm{con}}$ develops a peak that grows, narrows, and drifts toward the same interval with increasing system size [Fig.~\ref{fig:thermodynamics}(c)]. The data for larger network sizes show the same trend and are presented in SM \cite{sm}. 

The Binder and susceptibility data admit a common one-parameter finite-size-scaling description with estimated critical exponents. Here, finite-size scaling serves as an internal consistency check rather than a strict extraction of universal critical exponents; the complete collapse analysis is provided in the SM. Together with the single-peaked histograms and initialization tests presented there, the convergent Binder meeting and sharpening susceptibility support a continuous transition and disfavor both a size-independent crossover and first-order coexistence over the accessible sizes. Physically, the transition is the continuous loss of the uniform compatible component of the connection. Whether collective correlations survive after this component disappears is addressed next.

\begin{figure*}[t]
\centering
\includegraphics[width= \textwidth]{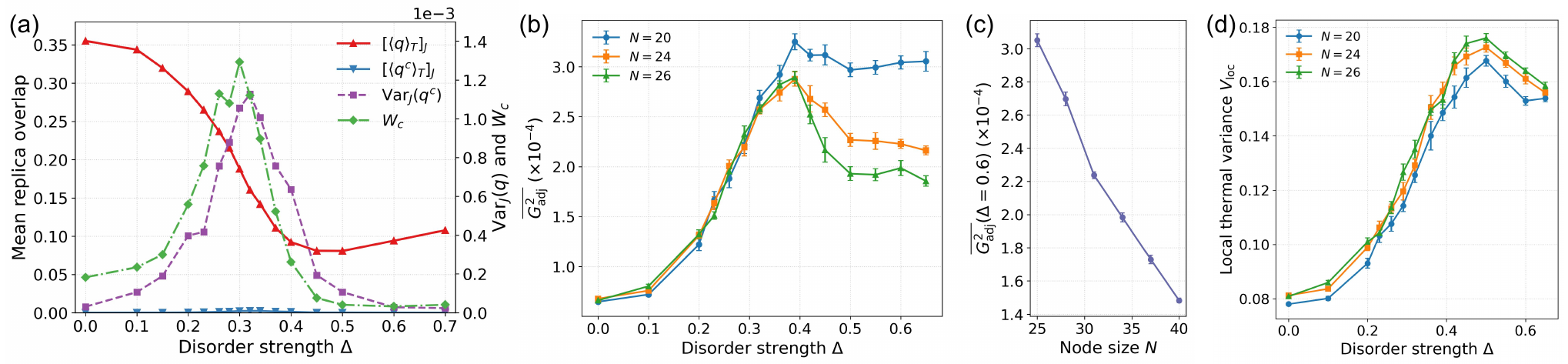}
\caption{Gauge-invariant characterization of the high-disorder phase. (a) Replica overlaps, width and sample-to-sample variance $\mathrm{Var}_J(q^c)$ for $N=30$. (b) $\overline{G_{\mathrm{adj}}^2}$ versus $\Delta$ for three sizes. (c) The finite-size analysis of $\overline{G_{\mathrm{adj}}^2}$ at representative $\Delta=0.6$. The approximate $N^{-1}$ scaling is consistent with $\mathcal{I}_{\rm adj}\sim 3N C_{\rm adj}$ remaining a finite value at $\Delta>\Delta_c$. (d) Local thermal variance of plaquettes rises sharply with disorder. Parameters used for $J_0$ and $\beta$ are identical to those in Figure~\ref{fig:thermodynamics}.}
\label{fig:correlated_liquid}
\end{figure*}

\emph{Correlated liquid instead of glass.---} 
Because two independent replicas respond to the same quenched one-point profile, their conventional overlap can remain nonzero even in the absence of glassy locking. We therefore test for replica-symmetry breaking using the connected overlap $q_{\alpha\gamma}^{\mathrm c}$ and its width $W_{\mathrm c}$, from which the complete disorder-conditioned one-point profile has been removed. In the high-disorder regime, $W_{\mathrm c}$ approaches the numerical noise floor, while $P(q_{\alpha\gamma}^{\mathrm c})$ narrows toward a single peak at the origin with increasing system size [Fig.~\ref{fig:correlated_liquid}(a) and SM]. The sample-to-sample fluctuations of the overlap statistic also decrease, indicating self-averaging over the accessible sizes. These finite-size trends provide no evidence for a thermodynamically persistent replica-overlap sector or replica-symmetry breaking \cite{Parisi1983,BinderYoung1986,Yucesoy2012,Billoire2014}, suggesting that the gauge network does not form glassy pattern even when highly frustrated.

The absence of replica locking does not imply that the disordered phase is featureless randomness. The integrated adjacent correlation weight $\mathcal I_{\mathrm{adj}}$ measures the total squared connected covariance carried by the adjacency neighborhood of a typical plaquette. Because replica subtraction removes the complete disorder-conditioned one-point profile, an ensemble of independently responding plaquettes has $\mathcal I_{\mathrm{adj}}=0$. Each triangular face has $z_{\mathrm{adj}}=3(N-3)$ adjacent faces, so the mean squared covariance of one adjacent pair is $\overline{G_{\mathrm{adj}}^2}=\frac{\mathcal I_{\mathrm{adj}}}{3(N-3)}$. The finite-size results show that $\overline{G_{\mathrm{adj}}^2}$ is enhanced towards the transition [Fig.~\ref{fig:correlated_liquid}(b)], and $\overline{G_{\mathrm{adj}}^2}\sim N^{-1}$ at a representative $\Delta>\Delta_c$ [Fig.~\ref{fig:correlated_liquid}(c)]. This renders $\mathcal I_{\mathrm{adj}}\sim N \overline{G_{\mathrm{adj}}^2}$ nearly constant with a finite large-$N$ limit. Thus the correlation of a particular adjacent pair becomes asymptotically weak, whereas the complete adjacent sector retains a finite integrated squared correlation per plaquette. This behavior is consistent with geometric dilution across the growing adjacency neighborhood.

In addition, the local thermal variance of plaquettes
$V_{\mathrm{loc}}=\left[F_N^{-1}\sum_f \left(\langle P_f^2\rangle_T-\langle P_f \rangle_T^2\right)\right]_J$
rises sharply through the transition region [Fig.~\ref{fig:correlated_liquid}(d)]. It means that disorder reduces the stiffness of local holonomy and broadens thermally reachable configurations, mimicking a dynamically fluctuating liquid rather than a frozen high-disorder configuration. The global disorder, the vanishing replica overlap width, the softened local stiffness, and spatially coordinated but geometrically diluted fluctuations altogether support a continuous transition from relatively stiff globally ordered phase to a thermally mobile, non-glassy correlated gauge in which collective correlations are distributed over an increasing number of neighboring simplices rather than concentrated in persistent pairwise locking.

\textit{Discussion and outlook.}--
In summary, we construct a thermodynamic theory of fluctuating relational transformations and the framework reveals an intricate way in which dense frustration organizes collective correlations without replica locking. The numerical results are consistent with the phase signature for a finite-size system
\begin{equation}
 M_P \ll 1,
 \quad \mathcal I_{\mathrm{adj}}\to
 \mathcal I_{\mathrm{adj}}^{\infty}>0,
 \quad W_{\mathrm c}\ll 1,
 \label{eq:phase-signature}
\end{equation}
with finite-size scaling suggesting that the conditions hold true as system size enlarges. These conditions separate the correlated gauge liquid from both an uncorrelated plaquette phase ($\mathcal I_{\mathrm{adj}}=0$) and a conventional spin-glass phase ($W_{\mathrm{c}}\sim\mathcal{O}(1)$). 

The coexistence of finite $\mathcal I_{\mathrm{adj}}$ and vanishing $W_{\mathrm c}$ follows directly from the network geometry. The adjacent sector contributes $W_{\mathrm{adj}}=\frac{\mathcal I_{\mathrm{adj}}}{\FN} =O(N^{-3})$ when $\mathcal I_{\mathrm{adj}}$ approaches a finite limit. It therefore carries a finite squared-correlation per plaquette while occupying a vanishing fraction of the $F_N^2$ face-pair space entering the global replica overlap. This geometric dilution allows dense frustration to sustain collective structure without macroscopic replica locking. The complete complex contains $O(N^3)$ plaquette constraints but only $O(N^2)$ link variables. These constraints are neither independent random bonds nor independent plaquette degrees of freedom: shared links and non-Abelian Bianchi identities organize them into geometrically constrained correlation sectors. The observed scaling therefore describes a distinct dense-network regime in which dense frustration can sustain an extensive adjacent correlation cloud without producing a macroscopic overlap manifold.

The result establishes a form of collective organization based on compatibility of relations rather than equality of node states, and promotes network compatibility to a many-body thermodynamic problem. In the communication interpretation, the near-flat phase admits mutually consistent translations between local frames, allowing global consensus. The gauge liquid phase does not allow a single globally compatible frame, yet local compatibilities retain a correlated and diffusing pattern after the static background is removed. The transition is therefore from global relational integrability to a correlated but non-glassy regime of path-dependent incompatibility, rather than simply from order to featureless disorder.

The construction in this work is not restricted to the complete complex or to $\SUtwo$. The separation of uniform coherence, disorder pinning, integrated loop correlations, and replica locking applies to group-valued relations on more general higher-order networks. The complete simplicial complex provides a controlled dense limit where these distinctions are especially sharp. Sparse, modular, and temporally evolving complexes may support additional regimes in which the range and support of the curvature correlations are controlled directly by network architecture. Determining this architecture-dependent phase structure and the full $(\beta,\Delta)$ phase diagram are natural directions for future work.

\begin{acknowledgments}
We thank Songlin Fang and Zhiqiang Yan for useful discussions on numerical computations of this work. This work is supported by the National Natural Science Foundation of China (NSFC) (Grant No. 12505042), and the NSFC of Zhejiang Province (Grant No. LQN25A050004).
\end{acknowledgments}

\bibliography{Aug16bib}
\appendix
\newpage
\section{Pinning, collective correlation, and glassiness}

For a fixed realization of the quenched couplings, define the equilibrium plaquette profile
\begin{equation}
 \mu_f^{(J)}=\avgT{P_f}.
 \label{eq:equilibrium_profile}
\end{equation}
Its uniform component is
\begin{equation}
 m_J=\frac1{\FN}\sum_f\mu_f^{(J)}=\avgT{m_P},
 \label{eq:sample_order_parameter}
\end{equation}
and the profile decomposes as
\begin{equation}
 \mu_f^{(J)}=m_J+\mu_{\perp,f}^{(J)},
 \qquad
 \sum_f\mu_{\perp,f}^{(J)}=0.
 \label{eq:profile_decomposition}
\end{equation}
The corresponding pinned weight is
\begin{align}
 Q_{\mathrm{pin}}^{(J)}
 &=\frac1{\FN}\sum_f\left(\mu_f^{(J)}\right)^2
 =m_J^2+Q_{\perp}^{(J)},
 \label{eq:pinned_decomposition}\\
 Q_{\perp}^{(J)}
 &=\frac1{\FN}\sum_f\left(\mu_{\perp,f}^{(J)}\right)^2.
 \label{eq:inhomogeneous_profile}
\end{align}
A finite $Q_{\perp}^{(J)}$ records a reproducible but nonuniform one-point profile selected by disorder. It does not by itself establish either collective fluctuations or glassiness.

Consider two equilibrium replicas, $\alpha$ and $\gamma$, sampled independently at the same quenched disorder. Their conventional plaquette overlap is
\begin{equation}
 Q_{\alpha\gamma}=\frac1{\FN}\sum_fP_f^\alpha P_f^\gamma.
 \label{eq:raw_replica_overlap}
\end{equation}
Conditional independence gives the exact identity
\begin{align}
 \avgT{Q_{\alpha\gamma}}
 &=\frac1{\FN}\sum_f
 \avgT{P_f^\alpha}\avgT{P_f^\gamma}
 \nonumber\\
 &=\frac1{\FN}\sum_f\left(\mu_f^{(J)}\right)^2
 =Q_{\mathrm{pin}}^{(J)}.
 \label{eq:raw_overlap_identity}
\end{align}
Thus a nonzero conventional overlap may be generated entirely by the common disorder-conditioned profile and is not, by itself, evidence of replica locking.

To remove this one-point background, define
\begin{equation}
 \delta P_f=P_f-\mu_f^{(J)},
 \qquad
 G_{fg}^{(J)}=\avgT{\delta P_f\delta P_g}.
 \label{eq:end_connected_covariance}
\end{equation}
Two distinct triangular faces are adjacent when they share an edge. Each face has
\begin{equation}
 z_{\mathrm{adj}}=3(N-3)
 \label{eq:end_adjacent_coordination}
\end{equation}
adjacent faces, and the number of unordered adjacent pairs is
\begin{equation}
 \Nadj
 =\EN\binom{N-2}{2}
 =\frac{\FN z_{\mathrm{adj}}}{2}.
 \label{eq:number_adjacent_pairs}
\end{equation}
The integrated adjacent correlation weight introduced in the Letter is
\begin{align}
 \mathcal I_{\mathrm{adj}}
 =\avgJ{\frac1{\FN}\sum_f\sum_{g\sim f}
 \left(G_{fg}^{(J)}\right)^2}
  =\avgJ{\frac2{\FN}\sum_{\langle f,g\rangle}
 \left(G_{fg}^{(J)}\right)^2},
 \label{eq:end_Iadj}
\end{align}
where the second sum is over unordered adjacent pairs. This normalization gives the average total squared connected covariance carried by the adjacency neighborhood of one plaquette.

The same quantity can be evaluated with two independent replicas. For $f\neq g$,
\begin{equation}
 \avgT{P_f^\alpha P_g^\gamma}
 =\mu_f^{(J)}\mu_g^{(J)},
 \label{eq:cross_replica_factorization}
\end{equation}
so that
\begin{align}
 \mathcal I_{\mathrm{adj}}
 =\avgJ{\frac2{\FN}\sum_{\langle f,g\rangle}
 \left(
 \avgT{P_f^\alpha P_g^\alpha}
 -\avgT{P_f^\alpha P_g^\gamma}
 \right)^2}.
 \label{eq:cross_replica_Iadj}
\end{align}
The subtraction removes both the uniform and inhomogeneous parts of the one-point profile. A matched independent-plaquette ensemble can reproduce every $\mu_f^{(J)}$ and every single-plaquette distribution, but it has $G_{fg}^{(J)}=0$ for $f\neq g$ and hence $\mathcal I_{\mathrm{adj}}=0$. 

The mean squared covariance of one randomly selected adjacent pair is not an independent order parameter; it is fixed by
\begin{equation}
 \overline{G_{\mathrm{adj}}^2}
 \equiv
 \avgJ{\frac1{\Nadj}\sum_{\langle f,g\rangle}
 \left(G_{fg}^{(J)}\right)^2}
 =\frac{\mathcal I_{\mathrm{adj}}}{z_{\mathrm{adj}}}
 =\frac{\mathcal I_{\mathrm{adj}}}{3(N-3)}.
 \label{eq:end_pairwise_dilution}
\end{equation}
Consequently, a finite thermodynamic limit of $\mathcal I_{\mathrm{adj}}$ implies the observed inverse-coordination scaling $\overline{G_{\mathrm{adj}}^2}\sim N^{-1}$.

Collective correlation does not necessarily imply glassy freezing. Define the connected replica overlap
\begin{equation}
 q_{\alpha\gamma}^{\mathrm c}
 =\frac1{\FN}\sum_f\delta P_f^\alpha\delta P_f^\gamma.
 \label{eq:connected_replica_overlap}
\end{equation}
Its mean vanishes for independent replicas, and conditional independence yields
\begin{equation}
 \avgT{\left(q_{\alpha\gamma}^{\mathrm c}\right)^2}
 =\frac1{\FN^2}\sum_{f,g}\left(G_{fg}^{(J)}\right)^2.
 \label{eq:connected_overlap_width}
\end{equation}
 After disorder averaging, decompose
\begin{equation}
 W_{\mathrm c}
 =W_{\mathrm{diag}}+W_{\mathrm{adj}}+W_{\mathrm{nonadj}}.
 \label{eq:end_sector_decomposition}
\end{equation}
The adjacent contribution is fixed exactly by
\begin{equation}
 W_{\mathrm{adj}}
 =\frac1{\FN^2}\avgJ{\sum_f\sum_{g\sim f}
 \left(G_{fg}^{(J)}\right)^2}
 =\frac{\mathcal I_{\mathrm{adj}}}{\FN}.
 \label{eq:end_Wadj}
\end{equation}
Thus $\mathcal I_{\mathrm{adj}}$ is the adjacent-sector contribution to the connected-overlap susceptibility $\FN W_{\mathrm c}$. If $\mathcal I_{\mathrm{adj}}\rightarrow\mathcal I_{\mathrm{adj}}^{\infty}<\infty$, then
\begin{equation}
 W_{\mathrm{adj}}=O(\FN^{-1})=O(N^{-3}).
 \label{eq:end_Wadj_scaling}
\end{equation}
The diagonal sector obeys
\begin{equation}
 0\leq W_{\mathrm{diag}}
 =\frac1{\FN^2}\avgJ{\sum_f\left(G_{ff}^{(J)}\right)^2}
 \leq\frac1{\FN}=O(N^{-3}),
 \label{eq:end_diagonal_bound}
\end{equation}
because $P_f\in[-1,1]$. Therefore, the finiteness of $\mathcal I_{\mathrm{adj}}$ and vanishing $W_c$ can coexist. Numerically, $W_{\mathrm c}$ is close to the noise floor in the high-disorder regime, while $P(q_{\alpha\gamma}^{\mathrm c})$ narrows with increasing $N$. The finite-size evolution of the overlap distribution and the unbiased estimator used for the squared adjacent correlator are reported in the Supplemental Material (SM). The no-glass conclusion is based on the observed decay of the total $W_{\mathrm c}$ and the narrowing of $P(q_{\alpha\gamma})$.

The phase distinction in the thermodynamic limit can therefore be succinctly summarized as
\begin{equation}
\begin{aligned}
&\lim_{N\to\infty}M_P(N,\Delta)=0,\\
&\lim_{N\to\infty}\mathcal I_{\mathrm{adj}}(N,\Delta)
=\mathcal I_{\mathrm{adj}}^\infty(\Delta)>0,\\
&\lim_{N\to\infty}W_{\mathrm c}(N,\Delta)=0.
 \end{aligned}
\label{eq:phase-signature}
\end{equation}
For $\Delta>\Delta_c$, the first condition removes global relational compatibility, the second excludes an independently responding plaquette liquid, and the third excludes a thermodynamically persistent glass sector. The simulations support this combination over the accessible sizes, with additional evidence presented in SM.

\clearpage
\appendix
\begin{widetext}
    
\section{\Large Supplemental Material}
\smallskip
This Supplemental Material documents the dense-network normalization, numerical implementation, frustration energy of the network, Monte Carlo estimators, finite-size-scaling analysis, energy-histogram check, and replica-overlap distributions used in the Letter. The exact decomposition of a conventional overlap into uniform and disorder-pinned one-point contributions, together with the combinatorial relation between the adjacent correlation sector and the global connected-overlap width, is given in the End Matter of the Letter and is not repeated in full here. 

\section{Geometry, normalization, and numerical representation}
\label{sec:S-geometry}

The complete simplicial $2$-complex on $N$ vertices contains
\begin{equation}
 \EN=\binom{N}{2},\qquad
 \FN=\binom{N}{3},\qquad
 c_N=N-2,
 \label{eq:S-counts}
\end{equation}
where $c_N$ is the number of triangular faces incident on one link. The identity
\begin{equation}
 3\FN=c_N\EN
 \label{eq:S-double-count}
\end{equation}
follows by counting face--edge incidences in two ways.

One matrix $U_{ij}\in\SUtwo$ is stored for every ordered pair $i,\ j$, with $U_{ji}=U_{ij}^{\dagger}$. For $i<j<k$ the oriented face variable and its normalized trace are
\begin{equation}
 U_{ijk}=U_{ij}U_{jk}U_{ik}^{\dagger},\qquad
 P_{ijk}=\frac12\operatorname{ReTr}U_{ijk}\in[-1,1].
 \label{eq:S-plaquette}
\end{equation}
The Hamiltonian and quenched couplings are exactly those of the Letter,
\begin{align}
 \mathcal H_J[U]&=-2\sum_{f=1}^{\FN}J_fP_f,
 \label{eq:S-Hamiltonian}\\
 J_f&=\frac{J_0}{c_N}+\frac{\Delta}{\sqrt{c_N}}\,\xi_f,
 \qquad \xi_f\sim\mathcal N(0,1).
 \label{eq:S-couplings}
\end{align}
The $c_N^{-1}$ and $c_N^{-1/2}$ factors are the dense-connectivity normalization adapted from the Sherrington--Kirkpatrick scaling \cite{SherringtonKirkpatrick1975}. The coherent field on a link is $c_N(J_0/c_N)=J_0$, while the root-mean-square random field is $\sqrt{c_N}(\Delta/\sqrt{c_N})=\Delta$. Both therefore remain finite as $N$ grows. Equation~\eqref{eq:S-double-count} also gives
\begin{equation}
 \mathcal H_0=-\frac{2J_0}{3}\EN m_P,
 \qquad
 m_P=\frac1{\FN}\sum_fP_f,
 \label{eq:S-uniform-energy}
\end{equation}
so the uniform contribution is extensive in the number of link variables $\EN$. For fixed $U$, the random part satisfies
\begin{equation}
 \operatorname{Var}_J(\mathcal H_{\rm rand}\mid U)
 =\frac{4\Delta^2}{c_N}\sum_fP_f^2=O(\EN).
 \label{eq:S-random-energy-scale}
\end{equation}
This establishes the same extensive scale used to normalize the susceptibility in the Letter.

An $\SUtwo$ matrix is represented by a unit quaternion
\begin{equation}
  q=(q_0,q_1,q_2,q_3),
  \qquad
  \sum_{\alpha=0}^{3}q_\alpha^2=1,
\end{equation}
through
\begin{equation}
  U(q)=q_0\mathbb I+i\sum_{a=1}^{3}q_a\sigma_a.
\end{equation}
Quaternion multiplication is
\begin{equation}
  (a_0,\bm a)(b_0,\bm b)
  =\left(a_0b_0-\bm a\cdot\bm b,
  a_0\bm b+b_0\bm a-\bm a\times\bm b\right),
  \label{eq:qmul}
\end{equation}
for the Pauli-matrix convention used in the code.  The conjugate is $(q_0,-\bm q)$ and
\begin{equation}
  \frac12\operatorname{ReTr}U(q)=q_0.
\end{equation}
The sign of the cross product in Eq.~\eqref{eq:qmul} depends on whether $U=q_0\mathbb I+i\bm q\cdot\bm\sigma$ or its complex-conjugate convention is used; the production code fixes one convention globally and is tested against explicit complex matrices.  Haar-random matrices are generated by drawing four independent normal variates and normalizing four independent Gaussian variables. A local proposal is generated as
\begin{equation}
  R=\cos\theta\,\mathbb I+i\sin\theta\,\hat{\bm n}\cdot\bm\sigma,
\end{equation}
with isotropic $\hat{\bm n}$ and a symmetric distribution of small rotation vectors.  The proposal $U_e' = RU_e$ is accepted with probability  \cite{Metropolis1953}
\begin{equation}
 p_{\rm acc}=\min\left[1,
 \exp\{-\beta(\mathcal H'_J-\mathcal H_J)\}\right]\,.
 \label{eq:S-metropolis}
\end{equation}

For every quenched realization $\{J_f\}$, two statistically independent parallel-tempering ensembles are evolved. The two ensembles share the same disorder but use independent initial states and Monte Carlo random numbers. This construction supplies the independent thermal replicas required by the connected-overlap and squared-covariance estimators.

Each ensemble contains configurations at inverse temperatures
\begin{equation}
 0<\beta_1<\beta_2<\cdots<\beta_{n_T}=\workbeta.
 \label{eq:S-beta-ladder}
\end{equation}
After local sweeps, exchanges are attempted between alternating neighboring temperature pairs. A proposed exchange between configurations with energies $\mathcal H_i$ and $\mathcal H_{i+1}$ is accepted with
\begin{equation}
 p_{i\leftrightarrow i+1}
 =\min\left\{1,
 \exp\left[(\beta_i-\beta_{i+1})(\mathcal H_i-\mathcal H_{i+1})\right]
 \right\},
 \label{eq:S-swap}
\end{equation}
which preserves the product equilibrium measure \cite{HukushimaNemoto1996}. Equilibration is assessed from several diagnostics: agreement between ordered and Haar-random starts where both are available; stationarity of $m_P$, overlap observables; and consistency of the two independent ensembles. Thermal averages are formed separately within each disorder realization before the outer disorder average. 

\section{Residual frustration energy}
\label{sec:S-frustration-energy}

The quenched couplings assign a preferred sign to every plaquette. If the plaquette variables could be chosen independently, the term $-2J_fP_f$ would be minimized by $P_f=\operatorname{sgn}(J_f)$. Thus, for each disorder realization $J\equiv\{J_f\}$,
\begin{equation}
 \mathcal H_{\rm ind}^{(J)}=-2\sum_f |J_f|
 \label{eq:S-independent-lower-bound}
\end{equation}
is a rigorous lower bound on the energy. In the simplicial gauge model this bound need not be attainable, because all plaquette holonomies are generated from shared link variables and must obey the non-Abelian compatibility constraints. We therefore define the residual frustration-energy gap
\begin{equation}
 \Delta\mathcal H_{\rm fr}^{(J)}
 =\avgT{\mathcal H_J}-\mathcal H_{\rm ind}^{(J)}\geq0.
 \label{eq:S-frustration-gap}
\end{equation}

At finite temperature, Eq.~\eqref{eq:S-frustration-gap} contains both genuine incompatibility of the quenched preferences and ordinary thermal excitation. Writing $E_0^{(J)}$ for the exact ground-state energy gives the decomposition
\begin{equation}
 \Delta\mathcal H_{\rm fr}^{(J)}
 =\left(E_0^{(J)}-\mathcal H_{\rm ind}^{(J)}\right)
 +\left(\avgT{\mathcal H_J}-E_0^{(J)}\right).
 \label{eq:S-frustration-decomposition}
\end{equation}
The first term is the zero-temperature frustration energy, whereas the second term is the thermal contribution. Since $E_0^{(J)}$ is not known separately for every sample, the simulations directly measure their sum in Eq.~\eqref{eq:S-frustration-gap}. We consequently refer to it as the \emph{residual} frustration energy.

We report two normalizations. The excess energy per fundamental link is
\begin{equation}
 \varepsilon_{\rm fr}
 =\avgJ{\frac{\Delta\mathcal H_{\rm fr}^{(J)}}{\EN}},
 \label{eq:S-frustration-per-link}
\end{equation}
while the dimensionless residual-frustration index is
\begin{align}
 r_{\rm fr}
 &=\avgJ{r_{\rm fr}^{(J)}},
 \nonumber\\
 r_{\rm fr}^{(J)}
 &=\frac{\Delta\mathcal H_{\rm fr}^{(J)}}
 {|\mathcal H_{\rm ind}^{(J)}|}
 =1-\frac{\avgT{\sum_fJ_fP_f}}{\sum_f|J_f|}.
 \label{eq:S-frustration-index}
\end{align}
The bound $|P_f|\leq1$ implies $0\leq r_{\rm fr}^{(J)}\leq2$. The value $r_{\rm fr}^{(J)}=0$ means that every locally available coupling energy is realized, whereas $r_{\rm fr}^{(J)}=1$ means that the weighted plaquette alignment $\avgT{\sum_fJ_fP_f}$ vanishes. The dimensionless index is the cleaner quantity for comparing different $N$ and $\Delta$, because the energy scale $\sum_f|J_f|$ itself changes with disorder. The quantity $\varepsilon_{\rm fr}$ displays the raw gap in units of the number of link degrees of freedom, but it is not guaranteed to approach an intensive limit: $\mathcal H_{\rm ind}^{(J)}$ is a deliberately formal independent-face bound, and $\sum_f|J_f|$ need not scale in the same way as $\EN$.

\begin{figure*}[t]
 \centering
 \begin{minipage}[t]{0.45\textwidth}
  \centering
  \textbf{(a)}\par\smallskip
  \includegraphics[width=\linewidth]
  {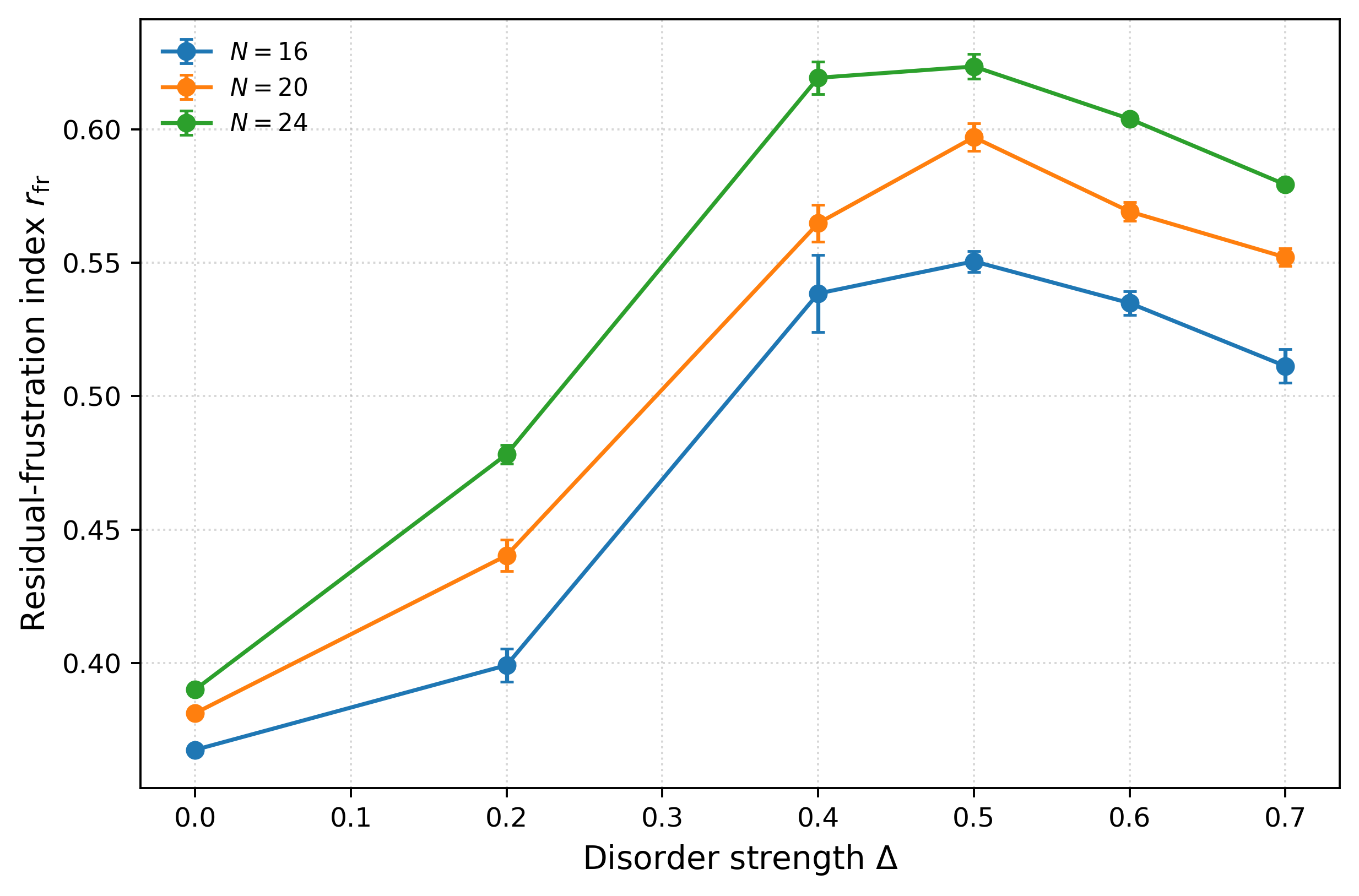}
 \end{minipage}\hfill
 \begin{minipage}[t]{0.45\textwidth}
  \centering
  \textbf{(b)}\par\smallskip
  \includegraphics[width=\linewidth]
  {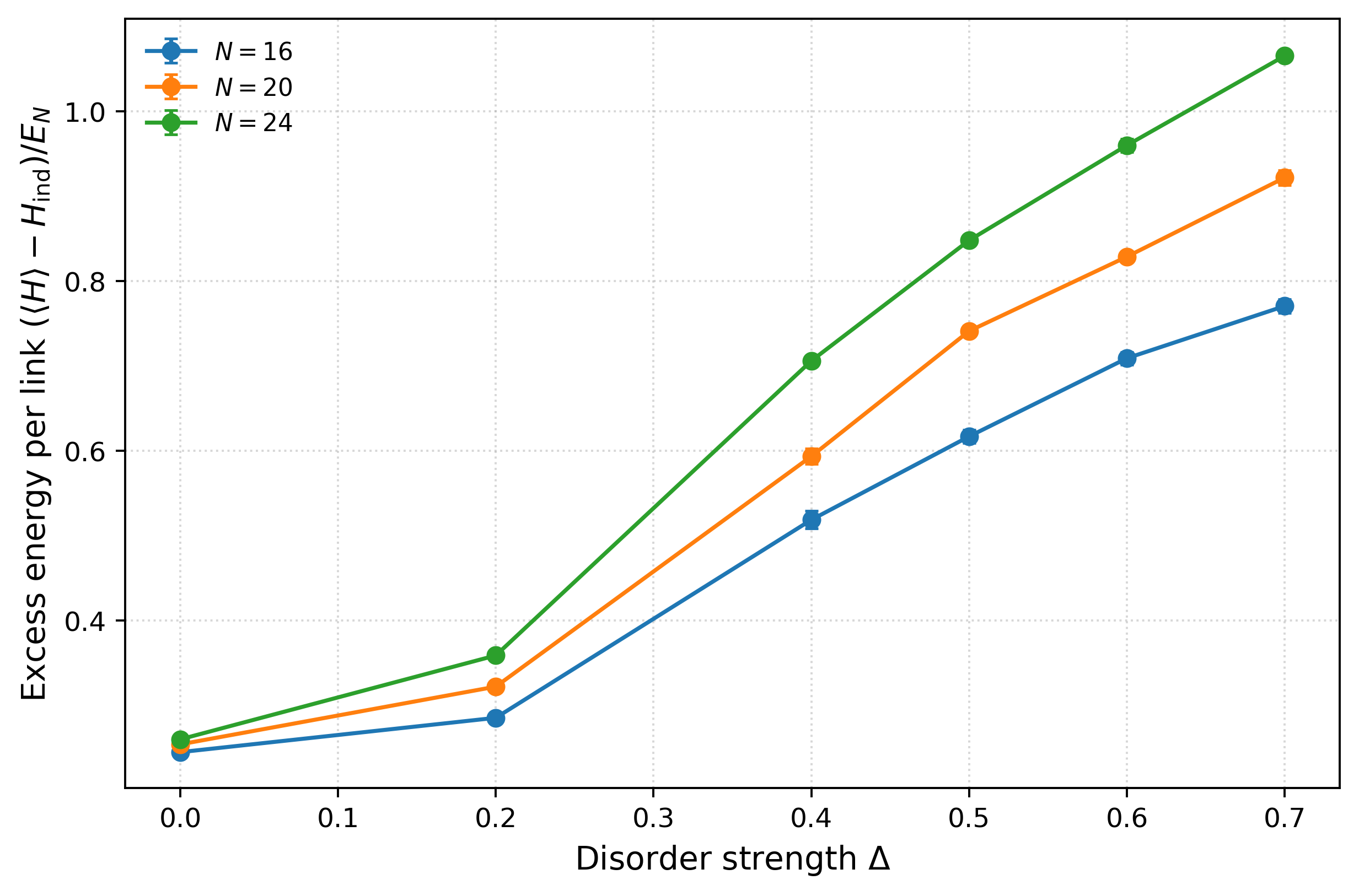}
 \end{minipage}
 \begin{minipage}[t]{0.45\textwidth}
  \centering
  \textbf{(c)}\par\smallskip
  \includegraphics[width=\linewidth]
  {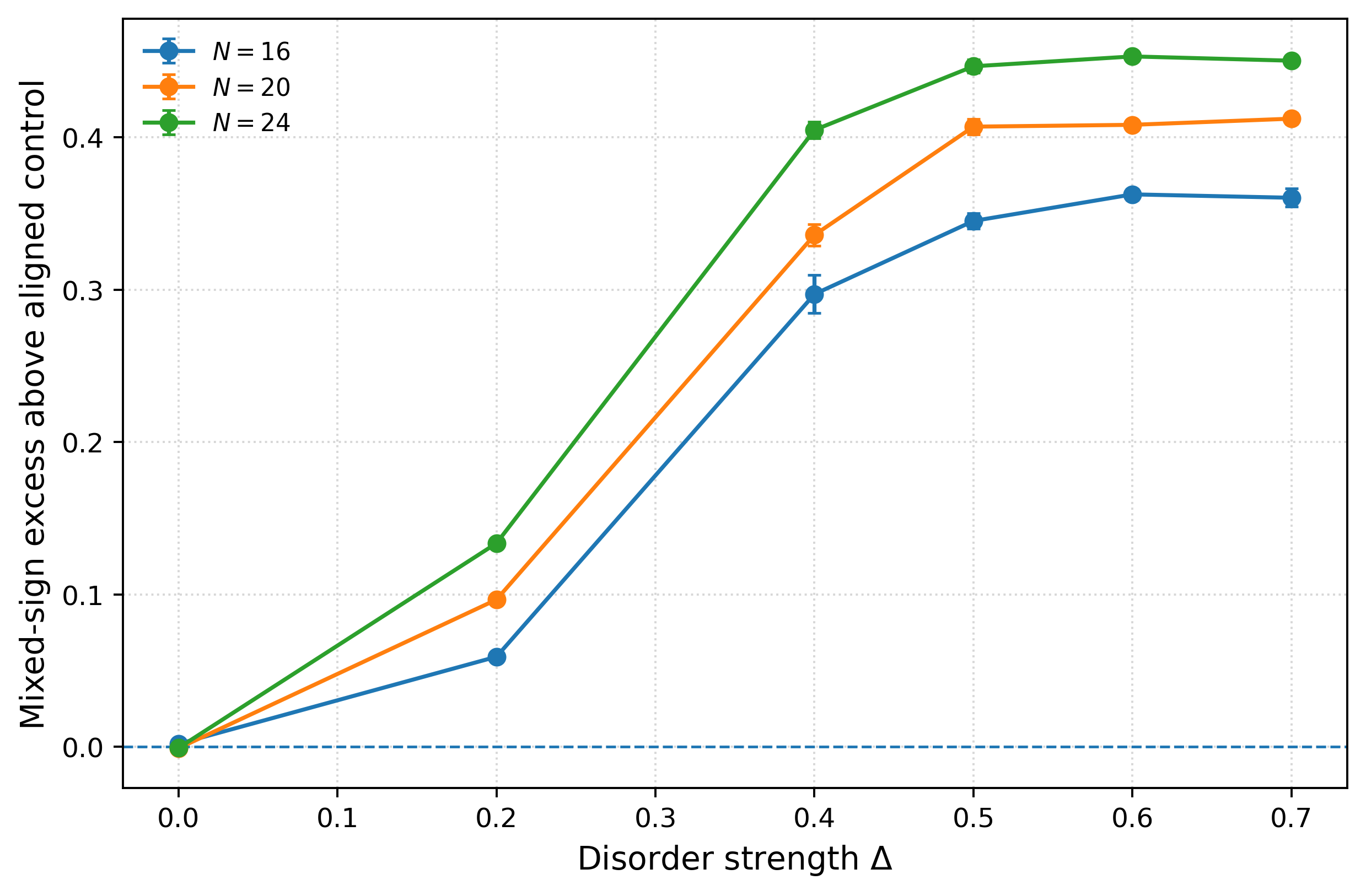}
 \end{minipage}
 \caption{Residual frustration diagnostics at $J_0=\workJzero$ and $\beta=\workbeta$. (a) Dimensionless residual-frustration index $r_{\rm fr}$ for the mixed-sign couplings. (b) Residual frustration energy  per link, $\varepsilon_{\rm fr}=(\avgT{\mathcal H_J}-\mathcal H_{\rm ind})/\EN$. (c) Matched excess $\Delta r_{\rm sign}$ obtained by subtracting the residual index of the sign-aligned control with the same $|J_f|$. Error bars are computed over independent disorder realizations. Since the finite-$\beta$ quantities include thermal excitation, panel (c) helps to assess the additional mismatch associated with incompatible coupling signs.}
 \label{fig:S-frustration-energy}
\end{figure*}

A matched sign-aligned control helps separate thermal mismatch from frustration caused by incompatible coupling signs. For every mixed-sign sample, we form a second sample with
\begin{equation}
 J_f^{(+)}=|J_f|.
 \label{eq:S-aligned-control}
\end{equation}
The two samples have identical coupling magnitudes. In the aligned sample, the configuration $U_{ij}=\mathbb I$ gives $P_f=1$ for every face and attains Eq.~\eqref{eq:S-independent-lower-bound} at zero temperature. We define the matched excess
\begin{equation}
 \Delta r_{\rm sign}
 =\avgJ{r_{\rm fr}^{(J)}-r_{\rm fr}^{(J,+)}},
 \label{eq:S-sign-frustration-excess}
\end{equation}
where $r_{\rm fr}^{(J,+)}$ is evaluated for the sign-aligned control. At $\Delta=0$ the mixed and aligned coupling sets coincide, so $\Delta r_{\rm sign}$ must be compatible with zero within uncertainty; this provides a useful implementation and equilibration check. A positive $\Delta r_{\rm sign}$ indicates an excess mismatch associated with the mixed coupling signs. This subtraction is a useful finite-temperature control, but it is not an exact projection onto the zero-temperature frustration energy because the mixed and aligned systems have different thermal spectra. The numerical results for the frustration energy are shown in Figure~\ref{fig:S-frustration-energy}.

The frustration observables are diagnostic of incompatible local preferences. In particular, a smooth evolution of $r_{\rm fr}$ through the critical region is compatible with a continuous loss of global compatibility. A large residual frustration energy does not imply glassiness: both a fluctuating liquid and a frozen glass can fail to realize the independent-plaquette lower bound. The transition order is determined from the Binder, susceptibility, scaling, and coexistence tests, whereas the liquid--glass distinction is determined from the profile-subtracted
connected-overlap sector.

\section{Thermodynamic observables and finite-size scaling}
\label{sec:S-FSS}
\begin{figure*}[t]
 \centering
 \includegraphics[width=0.6\textwidth]{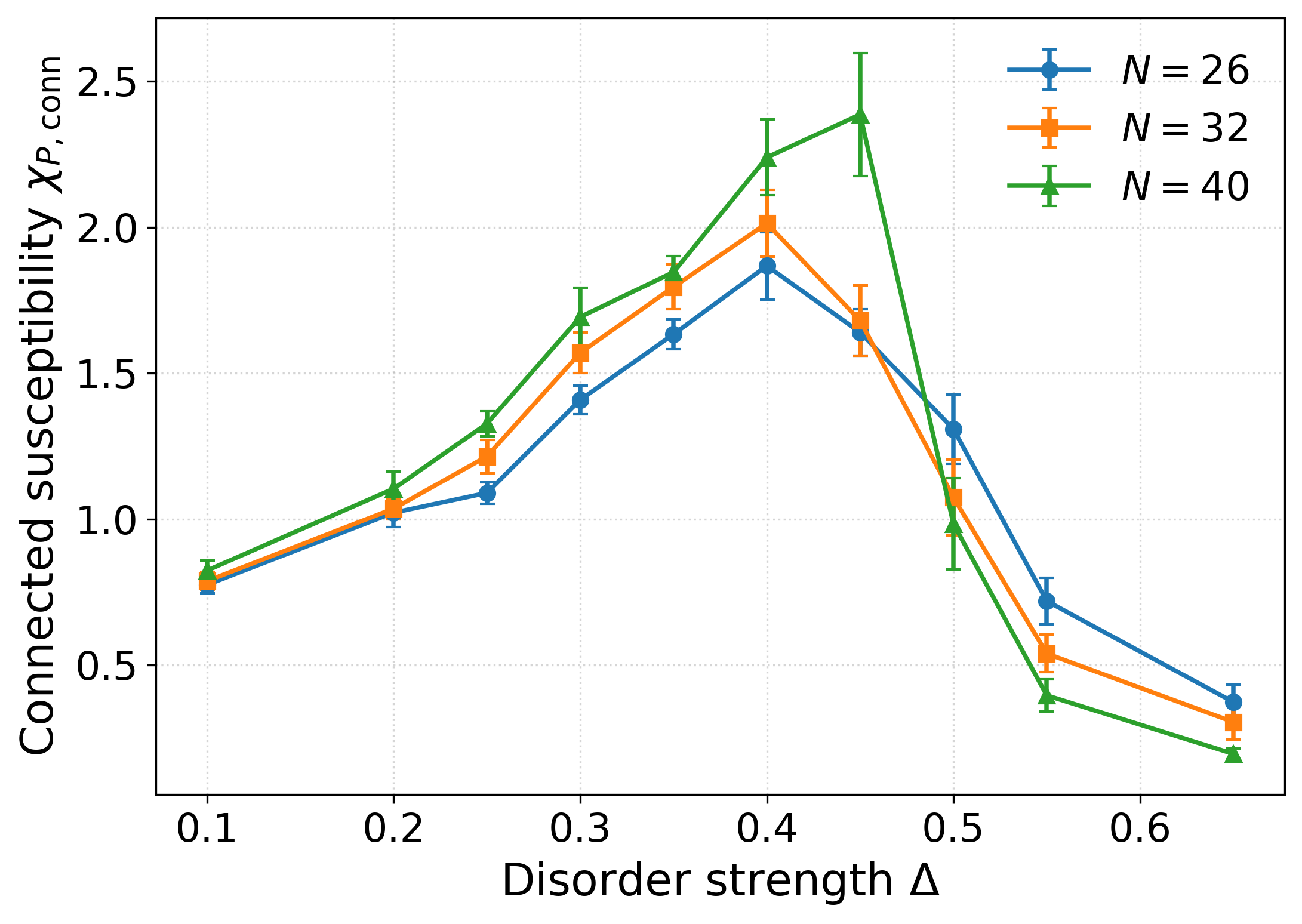}%
 \caption{Connected thermal susceptibility of the plaquettes $\chi_{P,\mathrm{con}}$ at larger node numbers show the same trend of a sub-extensive growth. The parameters used are the same as those in Figure 1(c) of the Letter. }
 \label{fig:chi}
\end{figure*}

The signed plaquette order parameter, Binder cumulant, and connected thermal susceptibility are
\begin{align}
 M_P&=\avgJ{\avgT{m_P}},
 \qquad m_P=\frac1{\FN}\sum_fP_f,
 \label{eq:S-MP}\\
 U_4&=1-\frac{\avgJ{\avgT{m_P^4}}}
 {3\,\avgJ{\avgT{m_P^2}}^2},
 \label{eq:S-U4}\\
 \chi_{\rm con}&=\beta\EN\,
 \avgJ{\avgT{m_P^2}-\avgT{m_P}^2}.
 \label{eq:S-chi}
\end{align}
No absolute value is taken in $M_P$: the positive mean coupling selects the near-flat sector, and the Hamiltonian has no $m_P\mapsto-m_P$ symmetry. The factor of three in $U_4$ is the Gaussian fourth-moment normalization. The factor $\EN$ in Eq.~\eqref{eq:S-chi} follows from the extensive scale in Eq.~\eqref{eq:S-uniform-energy}. The results for connected susceptibility are presented in Figure~\ref{fig:chi}.

\begin{figure*}[t]
 \centering
   \includegraphics[width=0.88\textwidth]{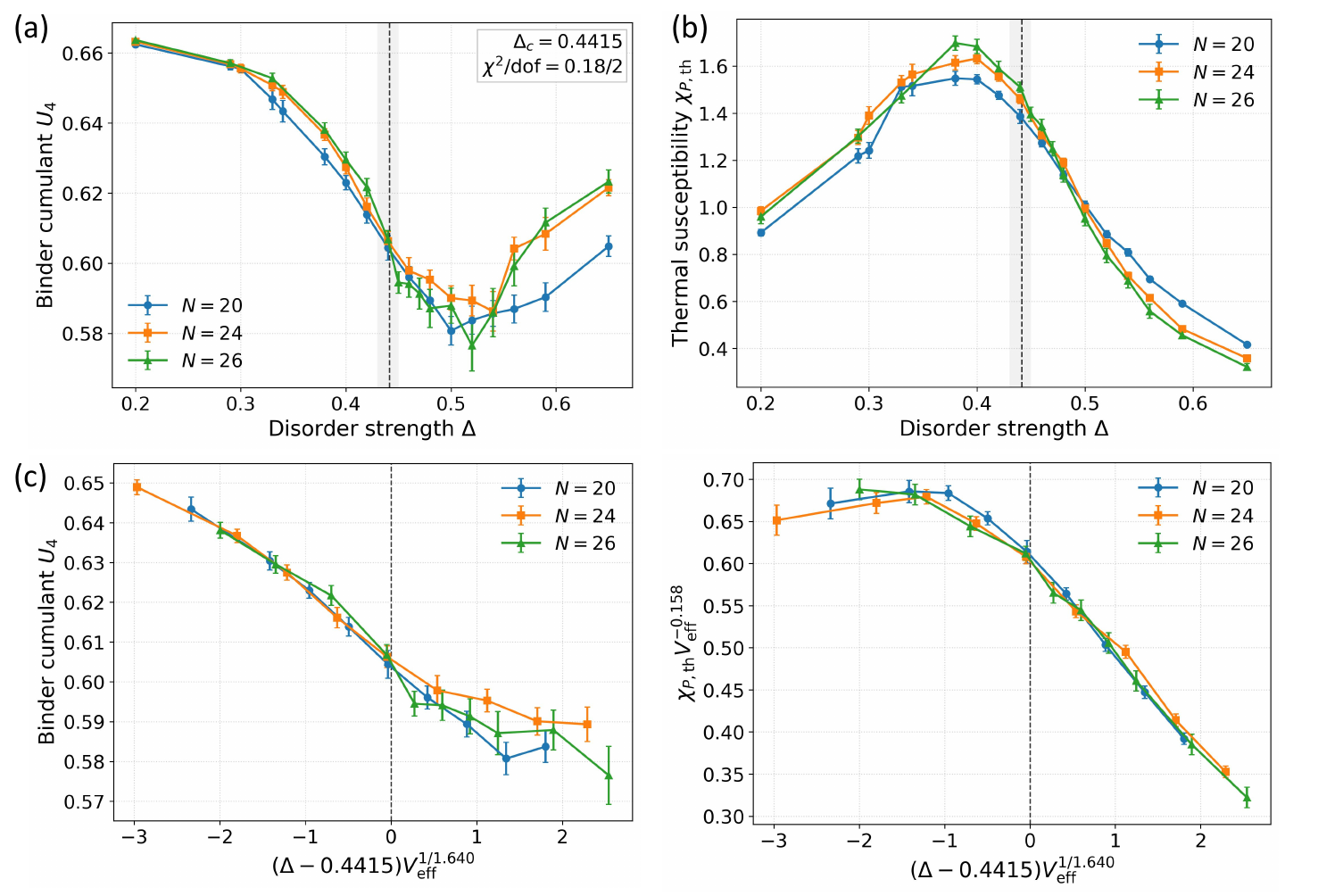}%
 \caption{Finite-size transition analysis from independent Monte Carlo runs. (a) The Binder cumulants for $N=20,24,26$ meet within the shaded interval $0.43\leq\Delta\leq0.45$; the vertical reference is $\Delta_c=0.44$. The small $\chi^2$ value supports a unique critical crossing. (b) The connected susceptibility develops a size-dependent peak in the same disorder region. (c) Binder-cumulant collapse using the variable $(\Delta-\Delta_c)\mathcal V_N^{1/\bar\nu}$. (d) Susceptibility collapse after rescaling by $\mathcal V_N^{\gamma_{\rm th}/\bar\nu}$. The displayed exponents, $\bar\nu\simeq 1.64$ and $\gamma_{\rm th}/\bar\nu\simeq0.16$, are effective diagnostic values rather than a universality-class determination.}
 \label{fig:S-FSS}
\end{figure*}

The complete complex has no Euclidean linear size. We therefore use the dense-network correlation volume
\begin{equation}
 \Veff=\frac{3\FN}{N}=\frac{(N-1)(N-2)}{2}
 \label{eq:S-Veff}
\end{equation}
and test the one-parameter forms
\begin{align}
 U_4(N,\Delta)&=\mathcal F_U\!\left[
 (\Delta-\Delta_c)\Veff^{1/\bar\nu}\right],
 \label{eq:S-U-collapse}\\
 \chi_{\rm con}(N,\Delta)&=
 \Veff^{\gamma_{\rm th}/\bar\nu}\,
 \mathcal F_\chi\!\left[
 (\Delta-\Delta_c)\Veff^{1/\bar\nu}\right].
 \label{eq:S-chi-collapse}
\end{align}
The selected transition data use $N=20,24,26$ [see Figure~\ref{fig:S-FSS}]. The direct Binder curves meet in the band $\Delta=\criticalBand$, and the reference value used for the collapse is $\Delta_c=\criticalDelta$. The collapse panels use the common interval $0.34\leq\Delta\leq0.52$. The joint diagnostic estimates are
\begin{equation}
 \bar\nu\simeq 1.64,
 \qquad
 \frac{\gamma_{\rm th}}{\bar\nu}\simeq 0.16,
 \label{eq:S-effective-exponents}
\end{equation}
with an estimated confidence interval $\bar\nu \in \left[1.2,2.6\right]$, and $\frac{\gamma_{\rm th}}{\bar\nu}\simeq 0.16\pm 0.03$. The accessible sizes strongly suggest a continuous transition with the above critical exponents. It is also compatible with the small $\chi^2$ value, which supports a unique critical crossing. But they do not justify assigning a universality class yet or interpreting these values as accurate asymptotic exponents.

A stable dimensionless-cumulant meeting, a growing and narrowing susceptibility, and a common scaling variable jointly disfavor a size-independent crossover. Conversely, the observed subextensive susceptibility growth over node numbers [see Figure~\ref{fig:chi}] and the absence of evident two-state structure disfavor the usual finite-size phenomenology of a first-order transition \cite{Binder1981,BinderLandau1984,ChallaLandauBinder1986}.

\begin{figure*}[t]
 \centering
   \includegraphics[width=0.72\textwidth]{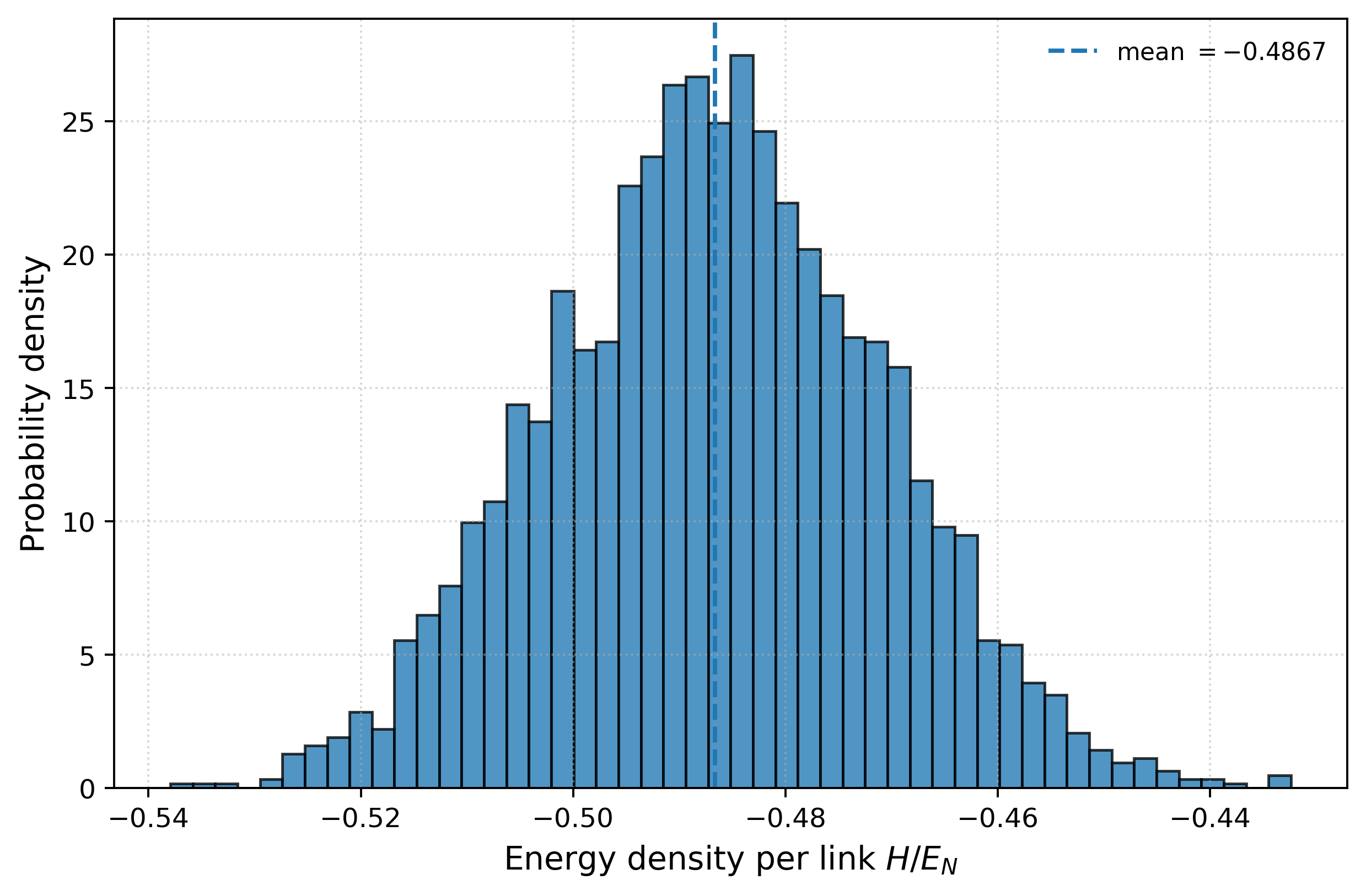}%
 \caption{Representative energy-density distribution $e=\mathcal H_J/E_N$  near the $N=20$ pseudocritical region at $J_0=1$, $\beta=6$, and $\Delta=0.40$. The distribution is single peaked within the resolution of the simulated history and shows no obvious two-state coexistence. Energy-density distribution near the $N=20$ pseudocritical region ($\Delta=0.40$). The single unimodal peak indicates the absence of phase coexistence over the accessible timescales.}
 \label{fig:S-histogram}
\end{figure*}

The energy density is normalized by the number of link variables,
\begin{equation}
 e=\frac{\mathcal H_J}{\EN}.
 \label{eq:S-energy-density}
\end{equation}
For a finite system, a clearly resolved two-peak energy distribution and persistent switching between the peaks would be standard evidence for phase coexistence. The histogram test is therefore useful as a check against an obvious first-order scenario, but a single unimodal histogram cannot by itself establish a continuous transition.

For $N=20$, a nearest simulated point to the size-dependent susceptibility maximum is $\Delta=0.40$, which need not coincide exactly with the thermodynamic estimate $\Delta_c\simeq0.44$. The corresponding long cold-replica history gives a single broad maximum rather than a resolved two-state structure. This observation is consistent with the Binder and susceptibility analysis, while remaining one component of the transition diagnosis. The energy histograms at other $\Delta$'s near the peak are also examined. The histograms are also consistent with the above conclusion (the diagrams are of shapes nearly identical to Fig.~\ref{fig:S-histogram}) and therefore we present Fig.~\ref{fig:S-histogram}) as a representative histogram.

A stable dimensionless-cumulant meeting, a growing and narrowing susceptibility, and a common scaling variable jointly disfavor a size-independent crossover. Conversely, the observed subextensive susceptibility growth over node numbers [see Figure~\ref{fig:chi}], the continuous global compatibility, and the absence of evident two-state structure in the singly-peaked energy histogram disfavor the usual finite-size phenomenology of a first-order transition \cite{Binder1981,BinderLandau1984,ChallaLandauBinder1986}. Therefore, the above evidence strongly suggests a continuous phase transition as disorder strength increases in the network. 

\section{Replica observables and finite-history estimators}
\label{sec:S-replicas}
At fixed disorder, let
\begin{equation}
 \mu_f^{(J)}=\avgT{P_f},\qquad
 \delta P_f=P_f-\mu_f^{(J)},\qquad
 G_{fg}^{(J)}=\avgT{\delta P_f\delta P_g}.
 \label{eq:S-profile-covariance}
\end{equation}
For two independent replicas $\alpha$ and $\gamma$, the unnormalized conventional overlap is
\begin{equation}
 Q_{\alpha\gamma}=\frac1{\FN}\sum_fP_f^\alpha P_f^\gamma.
 \label{eq:S-Qraw}
\end{equation}
Conditional independence gives
\begin{equation}
 \avgT{Q_{\alpha\gamma}}
 =\frac1{\FN}\sum_f\left(\mu_f^{(J)}\right)^2
 \equiv Q_{\rm pin}^{(J)}.
 \label{eq:S-pinning-identity}
\end{equation}
Thus a nonzero raw overlap can be generated entirely by the common quenched one-point profile. The orthogonal decomposition of $Q_{\rm pin}^{(J)}$ is derived in the End Matter.

Glassy locking is tested after subtracting this profile,
\begin{equation}
 q_{\alpha\gamma}^{\rm c}
 =\frac1{\FN}\sum_f\delta P_f^\alpha\delta P_f^\gamma,
 \qquad
 W_{\rm c}=\avgJ{\avgT{(q_{\alpha\gamma}^{\rm c})^2}}.
 \label{eq:S-qc-Wc}
\end{equation}
The thermal mean of $q_{\alpha\gamma}^{\rm c}$ vanishes for independent replicas, while its width obeys
\begin{equation}
 W_{\rm c}=\frac1{\FN^2}\avgJ{
 \sum_{f,g}\left(G_{fg}^{(J)}\right)^2}.
 \label{eq:S-Wc-covariance}
\end{equation}
A thermodynamically persistent glass sector would require a nonvanishing limiting width or an equivalently nontrivial limiting connected-overlap distribution. The finite-size narrowing of overlap distributions is therefore used in the same spirit as standard overlap-distribution diagnostics in spin glasses \cite{EdwardsAnderson1975,Yucesoy2012,Billoire2014}.

The bounded cosine overlap used for visualization is
\begin{equation}
 q_{\alpha\gamma}
 =\frac{\sum_fP_f^\alpha P_f^\gamma}
 {\sqrt{\sum_f(P_f^\alpha)^2\sum_f(P_f^\gamma)^2}}.
 \label{eq:S-cosine-overlap}
\end{equation}
Because its denominator fluctuates, Eq.~\eqref{eq:S-pinning-identity} is exact for $Q_{\alpha\gamma}$, not for the ratio in Eq.~\eqref{eq:S-cosine-overlap}. The raw cosine distribution may therefore be centered away from zero without implying a glass. The no-glass inference relies on the profile-subtracted width $W_{\rm c}$ and on the narrowing connected sector, not on the location of the raw-overlap mean.

For the adjacent covariance, an especially simple independent-replica product construction is available because the two thermal ensembles are independent. With $M$ measurements in each ensemble, define the sample covariances
\begin{equation}
 \widehat G_{fg}^{(r)}
 =\frac1{M-1}\sum_{t=1}^{M}
 \left(P_f^{r}(t)-\overline P_f^{r}\right)
 \left(P_g^{r}(t)-\overline P_g^{r}\right),
 \qquad r\in\{\alpha,\gamma\}.
 \label{eq:S-sample-covariance}
\end{equation}
Independence of the two histories gives
\[
\mathbb E[\widehat G_{fg}^{(\alpha)}\widehat G_{fg}^{(\gamma)}]
=\mathbb E[\widehat G_{fg}^{(\alpha)}]\,
 \mathbb E[\widehat G_{fg}^{(\gamma)}].
\]
After autocorrelation-aware blocking, each factor is a consistent estimate of $G_{fg}^{(J)}$; if the blocked covariance estimates are unbiased, their product is exactly unbiased for $(G_{fg}^{(J)})^2$. The pair-averaged estimator used in the adjacent-correlation calculation is therefore
\begin{equation}
 \widehat{\overline{G_{\rm adj}^2}}^{(J)}
 =\frac1{\Nadj}\sum_{\langle f,g\rangle}
 \widehat G_{fg}^{(\alpha)}\widehat G_{fg}^{(\gamma)}.
 \label{eq:S-unbiased-adjacent}
\end{equation}
Most importantly, this avoids the strictly positive square-of-noise bias that would arise from squaring one covariance estimate from a single history. The same two-ensemble histories give the local thermal variance through
\begin{equation}
 \widehat V_{\rm loc}^{(J)}
 =\frac1{2\FN}\sum_f
 \left[\widehat{\operatorname{Var}}_\alpha(P_f)
 +\widehat{\operatorname{Var}}_\gamma(P_f)\right].
 \label{eq:S-local-variance-estimator}
\end{equation}

\begin{figure*}[t]
 \centering
 \begin{minipage}[t]{0.9\textwidth}
  \centering
   \includegraphics[width=0.85\linewidth]{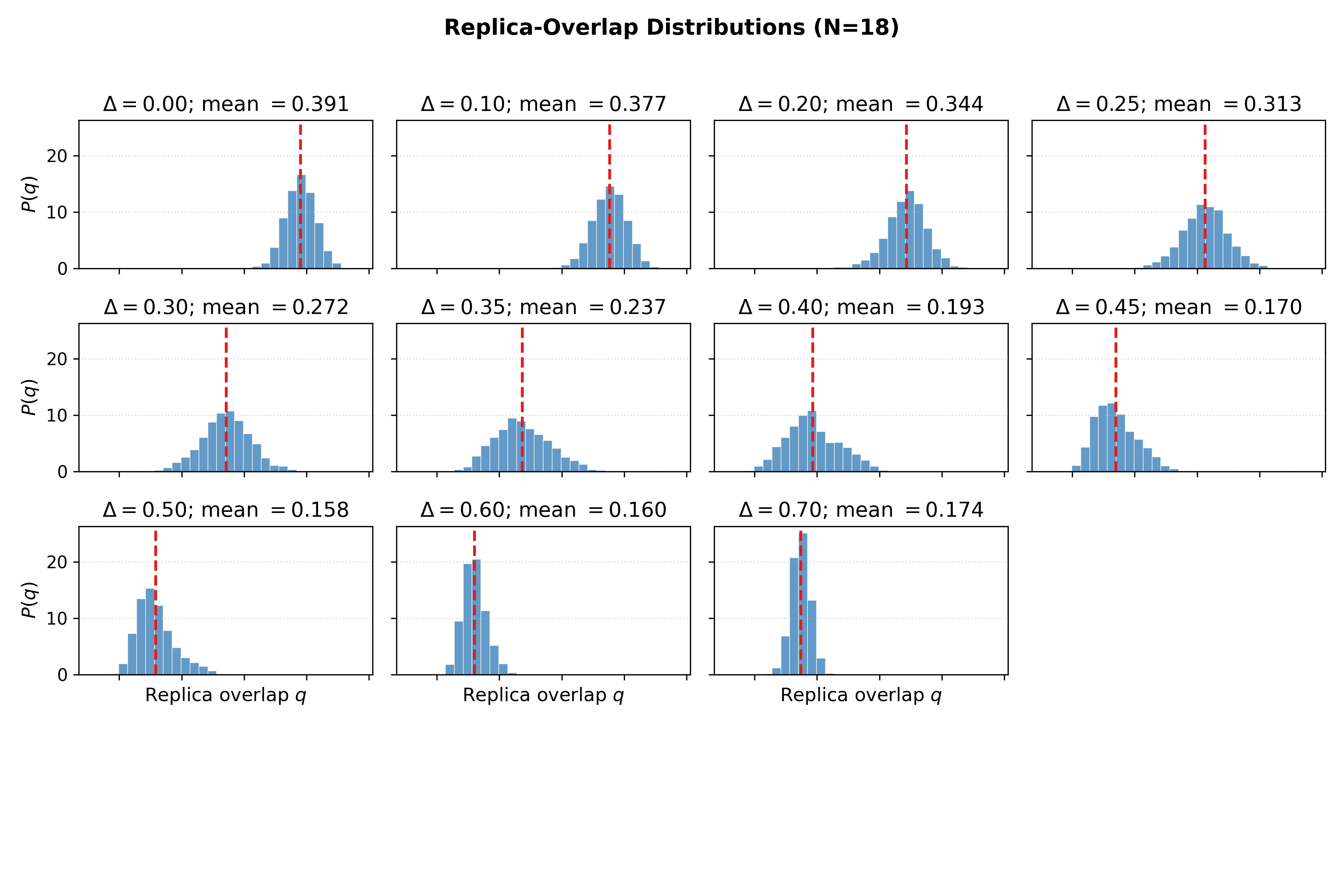}
   \end{minipage}\\
 \begin{minipage}[t]{0.85\textwidth}
  \centering
   \includegraphics[width=0.9\linewidth]{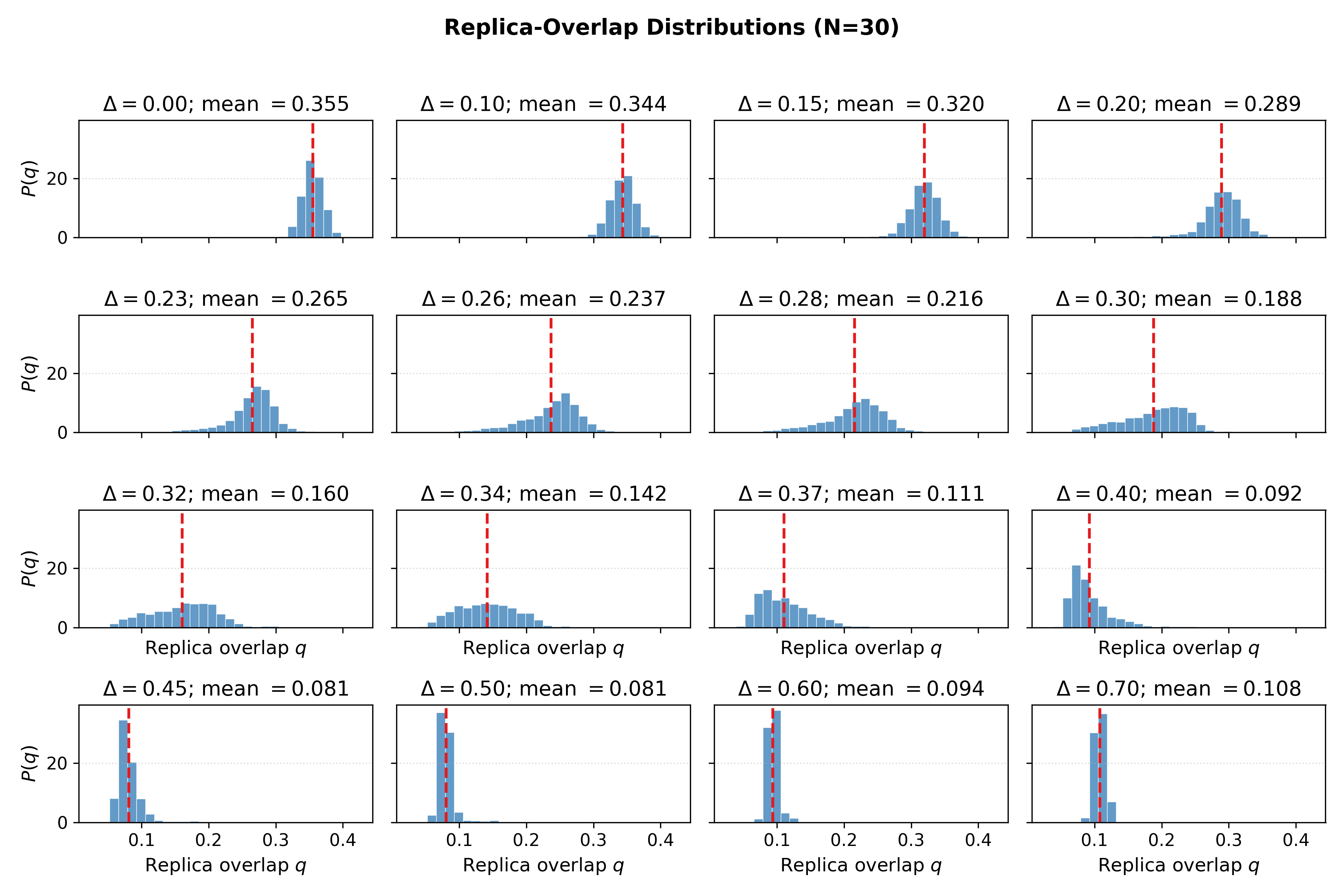}
 \end{minipage}
 \caption{Finite-size evolution of the replica-overlap distributions for (upper) $N=18$ and (lower) $N=30$, at the disorder values indicated in the panels. The raw overlap can remain centered at a nonzero value because both replicas share the same quenched one-point profile. Relative to $N=18$, the $N=30$ distributions are narrower and do not develop a persistent broad multi-state structure in the high disorder regime, for instance, at $\Delta=0.6$.}
 \label{fig:S-overlap-distributions}
\end{figure*}

\section{Integrated adjacent correlation and the no-glass consistency check}
\label{sec:S-adjacent}
Each triangular face has
\begin{equation}
 z_{\rm adj}=3(N-3)
 \label{eq:S-zadj}
\end{equation}
adjacent faces, and the number of unordered adjacent pairs is
\begin{equation}
 \Nadj=\EN\binom{N-2}{2}=\frac{\FN z_{\rm adj}}{2}.
 \label{eq:S-Nadj}
\end{equation}
The mean squared covariance of one adjacent pair and the integrated adjacent weight used in the Letter are related exactly by
\begin{align}
 \overline{G_{\rm adj}^2}
 &=\avgJ{\frac1{\Nadj}\sum_{\langle f,g\rangle}
 \left(G_{fg}^{(J)}\right)^2},
 \label{eq:S-pairwise-adjacent}\\
 \mathcal I_{\rm adj}
 &=\avgJ{\frac1{\FN}\sum_f\sum_{g\sim f}
 \left(G_{fg}^{(J)}\right)^2}
 =z_{\rm adj}\,\overline{G_{\rm adj}^2}.
 \label{eq:S-Iadj}
\end{align}
The numerical finite-size trend $\overline{G_{\rm adj}^2}\sim N^{-1}$ is therefore compatible with a finite integrated weight $\mathcal I_{\rm adj}$ over the accessible high-disorder sizes. The individual pair correlation is geometrically diluted because each face has $O(N)$ adjacent neighbors, while the total adjacent correlation carried by its neighborhood remains $O(1)$.

The adjacent contribution to the connected-overlap width is
\begin{equation}
 W_{\rm adj}
 =\frac1{\FN^2}\avgJ{\sum_f\sum_{g\sim f}
 \left(G_{fg}^{(J)}\right)^2}
 =\frac{\mathcal I_{\rm adj}}{\FN}.
 \label{eq:S-Wadj}
\end{equation}
Thus a finite $\mathcal I_{\rm adj}$ gives $W_{\rm adj}=O(N^{-3})$, fully consistent with a vanishing global width $W_{\rm c}$. The diagonal contribution also obeys $0\leq W_{\rm diag}\leq1/\FN$. All three sectors in
\begin{equation}
 W_{\rm c}=W_{\rm diag}+W_{\rm adj}+W_{\rm nonadj}
 \label{eq:S-sector-decomposition}
\end{equation}
are nonnegative. Consequently, the measured decay of the total $W_{\rm c}$ and a finite $\mathcal I_{\rm adj}$ are compatible. 

The combined finite-size signature, which are all supported by the our numerical results, is therefore
\begin{equation}
 M_P\to0,\quad
 \mathcal I_{\rm adj}\ \text{consistent with a finite positive limit},
 \quad  W_{\rm c}\to 0,\quad \text{and} \quad P\left(q_{\alpha\gamma}\right)
\longrightarrow \delta\left(q_{\alpha\gamma}-\avgJ{\avgT{q_{\alpha\gamma}}}\right)
 \label{eq:S-phase-signature}
\end{equation}
within the accessible sizes. The first trend is the loss of global compatibility, the second distinguishes the state from independently fluctuating plaquettes, and the third and forth provides no evidence for a thermodynamically persistent glass sector. The state is thus operationally defined as a correlated gauge liquid by the joint conditions in Eq.~\eqref{eq:S-summary-phase-signature}. 

Figure~\ref{fig:S-overlap-distributions} compares the archived overlap distributions for $N=18$ and $N=30$. The nonzero center of the raw overlap is permitted by Eq.~\eqref{eq:S-pinning-identity}: two independent replicas respond to the same disorder-conditioned face profile and are consistent with the mean values of $\langle q \rangle_T$. The connected overlap $q^c$ narrows towards $P\left(q_{\alpha\gamma}^{\mathrm{c}}\right)
\longrightarrow \delta\left(q_{\alpha\gamma}^{\mathrm{c}}\right)$ while the raw overlap $P\left(q_{\alpha\gamma}\right)
\longrightarrow \delta\left(q_{\alpha\gamma}-\avgJ{\avgT{q_{\alpha\gamma}}}\right)$. The useful finite-size information is the absence of a persistent broad or multi-peaked structure and the reduction of the width at the larger size. The narrowing replica overlap distributions and the vanishing $W_{\rm c}$ jointly demonstrate the conservation of replica symmetry in the high-disorder regime. Therefore, the results exclude the emergence of a glass phase even when the system is highly frustrated in its plaquette configuration.

\section{Summary of transition and phase diagnostics}
\label{sec:S-diagnostic-summary}

\begin{table}
\caption{Finite-size diagnostics of the transition.}
\label{tab:S-transition-diagnostics}
\centering
\scriptsize
\setlength{\tabcolsep}{5pt}
\renewcommand{\arraystretch}{1.25}
\resizebox{0.9\linewidth}{!}{%
\begin{tabular}{@{}lccc@{}}
\toprule
Observable & Continuous transition & Crossover & First order \\
\midrule
$M_P$ & Sharpens with $N$ & Width saturates & Jump / coexistence \\
$U_4$& Common crossing& No stable crossing& Strong anomaly \\
$\chi_{\rm con}$& Peak grows subextensively& Peak saturates& $\chi_{\rm con}^{\max}\propto E_N$ \\
FSS collapse& Stable& Unstable / absent& Fails \\
Energy histograms& Single peak& Single peak& Double peaks \\
\bottomrule
\end{tabular}%
}
\end{table}

\begin{table}[b]
\caption{Thermodynamic phase classification in the large-$N$ limit.}
\label{tab:S-phase-diagnostics}
\centering
\scriptsize
\setlength{\tabcolsep}{7pt}
\renewcommand{\arraystretch}{1.3}
\resizebox{0.8\linewidth}{!}{%
\begin{tabular}{@{}lcccc@{}}
\toprule
Phase
& $M_P$ & $\mathcal I_{\rm adj}$ & $W_{\rm c}$ & Character \\
\midrule
Coherent& $>0$ & --- & $\to0$ & Globally compatible \\
Independent liquid& $\to 0$ & $\to 0$ & $\to 0$ & Uncorrelated \\
Correlated gauge liquid& $\to0$& $>0$& $\to0$& Correlated, non-glassy \\
Gauge glass& $\to0$& ---& $>0$& Replica locked \\
\bottomrule
\end{tabular}%
}
\end{table}

We use two sets of diagnostics. The first determines whether the loss of global compatibility is a crossover, a first-order transition, or a continuous transition. The criteria are summarized in Table~\ref{tab:S-transition-diagnostics}. The second identifies the resulting equilibrium phase and is summarized in Table~\ref{tab:S-phase-diagnostics}. No single finite-size observable is decisive; the classification is based on their combined scaling behavior \cite{Binder1981,BinderLandau1984,ChallaLandauBinder1986}.

For the present data, the Binder curves meet near $\Delta_c\simeq0.44$, the susceptibility peak grows and narrows subextensively, and the data admit a common finite-size collapse. The energy histogram is single peaked and coherent and random initial states converge after equilibration. This combination supports a continuous transition rather than a size-independent crossover or a first-order transition.

The simulations support
\begin{equation}
 M_P\to0, \qquad  \mathcal I_{\mathrm{adj}} \to\mathcal I_{\mathrm{adj}}^\infty>0,
 \qquad  W_{\mathrm c}\to0,
 \label{eq:S-summary-phase-signature}
\end{equation}
which identifies the high-disorder state as a correlated gauge liquid.

The conventional replica overlap and the pinned weight $Q_{\mathrm{pin}}$ are not glass diagnostics, because both can remain nonzero solely through the disorder-conditioned one-point profile. Likewise, the residual frustration energy and the local thermal variance provide supporting information about incompatibility and mobility, but they do not by themselves determine the transition order or distinguish a liquid from a glass. The liquid--glass distinction is controlled by the large-$N$ behavior of $W_{\mathrm c}$ and the connected-overlap distribution \cite{Yucesoy2012,Billoire2014}.


\end{widetext}

\end{document}